\documentclass[aps,prc,preprint,groupedaddress,showpacs]{revtex4}
\usepackage[dvips]{graphicx}
\usepackage{amsmath,amssymb}

\begin{document}
\title{Instability of nuclear wobbling motion and tilted axis rotation}

\author{Masayuki Matsuzaki}
\email[]{matsuza@fukuoka-edu.ac.jp}
\author{Shin-Ichi Ohtsubo}
\affiliation{Department of Physics, Fukuoka University of Education, 
             Munakata, Fukuoka 811-4192, Japan}

\date{\today}

\begin{abstract}
   We study a possible correspondence between the softening of the 
wobbling mode and the ``phase transition" of the one-dimensionally rotating mean 
field to a three-dimensionally rotating one by comparing the properties of the 
wobbling mode obtained by the one-dimensional cranking model $+$ random phase 
approximation with the total routhian surface obtained by the three-dimensional
tilted-axis cranking model. The potential surface for the observed wobbling 
mode excited on the triaxial superdeformed states in $^{163}$Lu is also analyzed. 
\end{abstract}

\maketitle

\section{Introduction}

 The concept of the phase transition of the mean field is useful for describing 
structure changes in the atomic nucleus although it is a quantum system composed 
of finite number of Fermions. Typical example is that a spherical mean field 
becomes unstable as the quadrupole vibration excited on top of it softens with 
changing particle numbers, then an axially symmetric mean field substitutes. 
This can rotate about one of the axes perpendicular to the symmetry axis. 
Consecutively, the axial symmetric mean field can become unstable as the 
$\gamma$ vibration softens, then a triaxially deformed mean field substitutes. 
This can rotate about all the three principal axes. Usually, however, a rotation 
about one axis dominates because the rotation about the axis with the largest 
moment of inertia is energetically favorable. When some excitation energy is 
supplied, small rotations about other two axes become possible. Consequently 
this produces a kind of vibrational motion of the rotational axis, that is, 
the wobbling motion. 

 The small amplitude wobbling motion at high spins was first 
discussed by Bohr and Mottelson~\cite{bm} in terms of a macroscopic 
rotor model. Then it was studied microscopically by 
Janssen and Mikhailov~\cite{jm} and Marshalek~\cite{ma} in terms of 
the random phase approximation (RPA). Since the small amplitude 
wobbling mode has the same quantum number, parity $\pi = +$ 
and signature $\alpha = 1$, as the odd-spin member of the 
$\gamma$ vibrational band, Mikhailov and Janssen~\cite{mj} 
anticipated that it would appear as a high-spin continuation 
of the odd-spin $\gamma$ band. But it has not been clear 
in which nuclei, at what spins, and with what shapes 
it would appear. Using the RPA, Shimizu and Matsuyanagi~\cite{sm} 
studied Er isotopes with small $|\gamma|$, Matsuzaki~\cite{mm} 
and Shimizu and Matsuzaki~\cite{smm} studied $^{182}$Os with a 
rather large negative $\gamma$ but their correspondence to 
the experimental data was not very clear. In 2001, 
{\O}deg{\aa}rd {\it et al.}~\cite{lu1} found an excited triaxial 
superdeformed band in $^{163}$Lu and identified it firmly as a 
wobbling band by comparing the observed and theoretical interband 
$E2$ transition rates. These data were investigated in terms of 
a particle-rotor model by Hamamoto~\cite{hama} and in terms of 
the RPA by Matsuzaki {\it et al.}~\cite{msmr}. In 2002, two-phonon 
wobbling excitations were also observed by Jensen 
{\it et al.}~\cite{2phonon} and their excitation energies show some 
anharmonicity. 

 The one-dimensionally rotating triaxial mean field may become unstable as the 
wobbling mode softens with changing some parameters. One of the present authors 
(MM) pointed out its theoretical possibility in the preceding Rapid 
Communication~\cite{msmr}. 
The possibility of this phase transition was discussed in terms of 
the harmonic oscillator model by Cuypers~\cite{cuy} and Heiss and 
Nazmitdinov~\cite{heiss} but their conclusions are controversial. 
A theoretical framework to describe three-dimensional rotations, 
possibly with large amplitude, was first devised by Kerman and 
Onishi~\cite{ko} within a time-dependent variational formalism. 
Onishi~\cite{oni} and Horibata and Onishi~\cite{hori} applied it 
to $^{166}$Er and $^{182}$Os, respectively. See Ref.~\cite{oi}, 
for example, for recent applications. The three-dimensional 
cranking model was first used by Frisk and 
Bengtsson~\cite{fb}. The word, ``tilted (axis) cranking (TAC)" was, 
to our knowledge, first used by Frauendorf~\cite{frau2} and 
it was applied to a kind of two-dimensional rotation --- the so-called 
shears band, observed for example in the $A \sim 200$ 
region~\cite{pb1,pb2,pb3}. Applications to multiquasiparticle 
high-$K$ bands were also extensively done; see Ref.~\cite{frau3} 
and references cited therein. When the rotation becomes fully 
three-dimensional, the new concept, chirality, 
emerges~\cite{chiral1,chiral2}. The tilted axis cranking was also applied 
to this~\cite{chiral3}. At finite temperature, the degree of freedom of 
spin orientation was studied macroscopically~\cite{hot1} and 
microscopically~\cite{hot2}. A relativistic formulation of the 
three-dimensional cranking was given by 
Kaneko {\it et al.}~\cite{kaneko} as an extension of the one-dimensional 
one given by the Munich group~\cite{koepf}. Madokoro {\it et al.}~\cite{mado} 
studied the shears band in $^{84}$Rb starting from the meson 
exchange interaction although the pairing field was neglected. 

 The purpose of the present paper is to elucidate the work in Ref.~\cite{msmr} 
by comparing two types of theoretical calculations, the one-dimensional 
cranking model $+$ RPA and the three-dimensional (tilted axis) cranking model. 
The former gives the ``mass parameters" for the motion of the angular momentum 
vector, that is, the moments of inertia, while the latter provides the surfaces 
on which the angular frequency vector moves around. 

\section{Model}

 We start from a one-body Hamiltonian in the rotating frame, 
\begin{gather}
h'=h-h_\mathrm{cr}, 
\label{hcr}\\
h=h_\mathrm{Nil}-\Delta_\tau (P_\tau^\dagger+P_\tau)
                   -\lambda_\tau N_\tau , \label{hsp} \\
h_\mathrm{Nil}=\frac{\mathbf{p}^2}{2M}
                +\frac{1}{2}M(\omega_x^2 x^2 + \omega_y^2 y^2 + \omega_z^2 z^2) \notag \\
                +v_{ls} \mathbf{l\cdot s} 
                +v_{ll} (\mathbf{l}^2 - \langle\mathbf{l}^2\rangle_{N_\mathrm{osc}}) .
                \label{hnil}
\end{gather}
($h_\mathrm{cr}$ is specified below.)
In Eq.(\ref{hsp}), $\tau = 1$ and 2 stand for neutron and proton, respectively, 
and chemical potentials $\lambda_\tau$ are determined so as to give correct average 
particle numbers $\langle N_\tau \rangle$. 
The oscillator frequencies in Eq.(\ref{hnil}) 
are related to the quadrupole deformation parameters $\epsilon_2$ and $\gamma$ 
in the usual way. (We adopt the so-called Lund convention.) 
They are treated as parameters as well as pairing gaps $\Delta_\tau$. 
The orbital angular momentum $\mathbf{l}$ in Eq.(\ref{hnil}) is defined in the 
singly-stretched coordinates $x_k' = \sqrt{\frac{\omega_k}{\omega_0}}x_k$, 
with $k =$ 1 -- 3 denoting $x$ -- $z$, 
and the corresponding momenta. 

\subsection{One-dimensional cranking model $+$ random phase approximation}

 Equations~(\ref{hcr}) -- (\ref{hnil}) with
\begin{equation}
h_\mathrm{cr}=\hbar\omega_\mathrm{rot}J_x 
\end{equation}
generate the system rotating one-dimensionally. 
Then, since $h'$ conserves parity $\pi$ and signature $\alpha$, nuclear states can be 
labeled by them. Nuclear states with quasiparticle (QP) excitations are obtained by 
exchanging the QP energy and wave functions such as 
\begin{equation}
(-e'_\mu, \mathbf{V}_\mu, \mathbf{U}_\mu) \rightarrow 
(e'_{\bar\mu}, \mathbf{U}_{\bar\mu}, \mathbf{V}_{\bar\mu}) ,
\label{exch}
\end{equation}
where ${\bar\mu}$ denotes the signature partner of $\mu$. 
We perform the RPA to the residual pairing plus 
doubly-stretched quadrupole-quadrupole ($Q'' \cdot Q''$) interaction between QPs. 
Since we are interested in the wobbling motion that has a definite quantum number, 
$\alpha = 1$, only two components out of five of the $Q'' \cdot Q''$ interaction 
are relevant. They are given by 
\begin{equation}
H_\mathrm{int}^{(-)}=-\frac{1}{2}\sum_{K=1,2} \kappa_K^{(-)} Q_K''^{(-)\dagger} Q_K''^{(-)} ,
\end{equation}
where the doubly-stretched quadrupole operators are defined by 
\begin{equation}
Q_K''=Q_K(x_k\rightarrow x_k'' = \frac{\omega_k}{\omega_0}x_k) ,
\end{equation}
and those with good signature are 
\begin{equation}
Q_K^{(\pm)}=\frac{1}{\sqrt{2(1+\delta_{K0})}}\left(Q_K \pm Q_{-K}\right) .
\end{equation}
The residual pairing interaction does not contribute because $P_\tau$ is an 
operator with $\alpha = 0$. 
The equation of motion, 
\begin{equation}
\left[h'+H_\mathrm{int}^{(-)},X_n^\dagger\right]_\mathrm{RPA}
=\hbar\omega_n X_n^\dagger ,
\end{equation}
for the eigenmode 
\begin{equation}
X_n^\dagger=\sum_{\mu<\nu}^{(\alpha=\pm 1/2)}
\Big(\psi_n(\mu\nu)a_\mu^\dagger a_\nu^\dagger 
+\varphi_n(\mu\nu)a_\nu a_\mu\Big)
\end{equation}
leads to a pair of coupled equations for the transition amplitudes 
\begin{equation}
T_{K,n}=\left\langle\left[Q_K^{(-)},X_n^\dagger\right]\right\rangle .
\end{equation}
Then, by assuming $\gamma \neq 0$, this can be cast~\cite{ma} into the form 
\begin{gather}
(\omega_n^2-\omega_\mathrm{rot}^2)\left[\omega_n^2-\omega_\mathrm{rot}^2
\frac{\left(\mathcal{J}_x-\mathcal{J}_y^\mathrm{(eff)}(\omega_n)\right)
   \left(\mathcal{J}_x-\mathcal{J}_z^\mathrm{(eff)}(\omega_n)\right)}
{\mathcal{J}_y^\mathrm{(eff)}(\omega_n)\mathcal{J}_z^\mathrm{(eff)}(\omega_n)}
                                  \right] \notag \\
=0 ,
\end{gather}
which is independent of $\kappa_K^{(-)}$s. 
This expression proves that the spurious mode ($\omega_n=\omega_\mathrm{rot}$; 
not a real intrinsic excitation but a rotation as a whole) given by the first 
factor and all normal modes given by the second are decoupled from each other. 
Here $\mathcal{J}_x = \langle J_x \rangle/\omega_\mathrm{rot}$ as usual and the 
detailed expressions of $\mathcal{J}_{y,z}^\mathrm{(eff)}(\omega_n)$ are given in 
Refs.~\cite{ma,mm,smm}. Among normal modes, one obtains 
\begin{equation}
\omega_\mathrm{wob}=\omega_\mathrm{rot}
\sqrt{\frac{\left(\mathcal{J}_x-\mathcal{J}_y^\mathrm{(eff)}(\omega_\mathrm{wob})\right)
   \left(\mathcal{J}_x-\mathcal{J}_z^\mathrm{(eff)}(\omega_\mathrm{wob})\right)}
     {\mathcal{J}_y^\mathrm{(eff)}(\omega_\mathrm{wob})
      \mathcal{J}_z^\mathrm{(eff)}(\omega_\mathrm{wob})}} ,
\label{disp}
\end{equation}
by putting $\omega_n=\omega_\mathrm{wob}$. 
Note that this gives a real excitation only when the argument of the square root 
is positive and it is non-trivial whether a collective solution appears or not. 
Evidently this coincides with the form derived by Bohr and Mottelson in a rotor 
model~\cite{bm} and known in classical mechanics~\cite{landau}. 

\subsection{Three-dimensional (tilted axis) cranking model}

In this model the one-body Hamiltonian is given by Eqs.(\ref{hcr}) -- (\ref{hnil}) with
\begin{gather}
h_\mathrm{cr}=\hbar\mathbf{\Omega}\cdot\mathbf{J}, \\
\mathbf{\Omega}=\omega_\mathrm{rot}
(\cos{\theta},\sin{\theta}\cos{\varphi},\sin{\theta}\sin{\varphi}).
\end{gather}
Pairing correlation is taken into the system by a simple BCS approximation with fixed gaps 
as in the case of the one-dimensional cranking. 
The expectation value $\langle\mathbf{J}\rangle$ calculated at each 
$(\omega_\mathrm{rot},\theta,\varphi)$ 
has three non-zero components in general; the stationary state that minimizes the total 
routhian is obtained by requiring 
$\langle\mathbf{J}\rangle\parallel\mathbf{\Omega}$ (see Ref.~\cite{frau3} for details). 
Obtained tilted solutions do not possess the signature symmetry and therefore describe 
$\Delta I = 1$ rotational bands. 
In the present work, given a set of mean-field parameters, $N_\tau$, $\epsilon_2$, 
$\gamma$, and $\Delta_\tau$, a configuration is specified at 
$\theta=0^\circ$ (principal axis cranking about the $x$ axis). 
Then by changing $\theta$ and $\varphi$ step by step, 
the most overlapped state is chased. This procedure gives an energy (total routhian) 
surface for the angular frequency vector. 
Surfaces for QP excited configurations can also be calculated by adopting a procedure 
similar to Eq.(\ref{exch}). 

\section{Result and discussion}

 For this first comparative calculation, we choose the 
$[(\nu h_{9/2},f_{7/2})^2(\pi h_{11/2})^2]_{16^+}$ four quasiparticle configuration in $^{146}$Gd 
among this mass region in which many oblate isomers have been observed. 
This state is described by $\epsilon_2 = 0.19$, $\gamma= 60^\circ$, 
$\Delta_n = 0.8$ MeV, $\Delta_p = 0.6$ MeV, and $\hbar\omega_\mathrm{rot} = 0.25$ MeV. 
Calculations are performed in the model space of three major shells; 
$N_\mathrm{osc} =$ 4 -- 6 for neutrons and 3 -- 5 for protons. 
The strengths of the $\mathbf{l\cdot s}$ and $\mathbf{l}^2$ potentials are 
taken from Ref.~\cite{br}. 

\begin{figure}[htbp]
  \includegraphics[width=7cm,keepaspectratio]{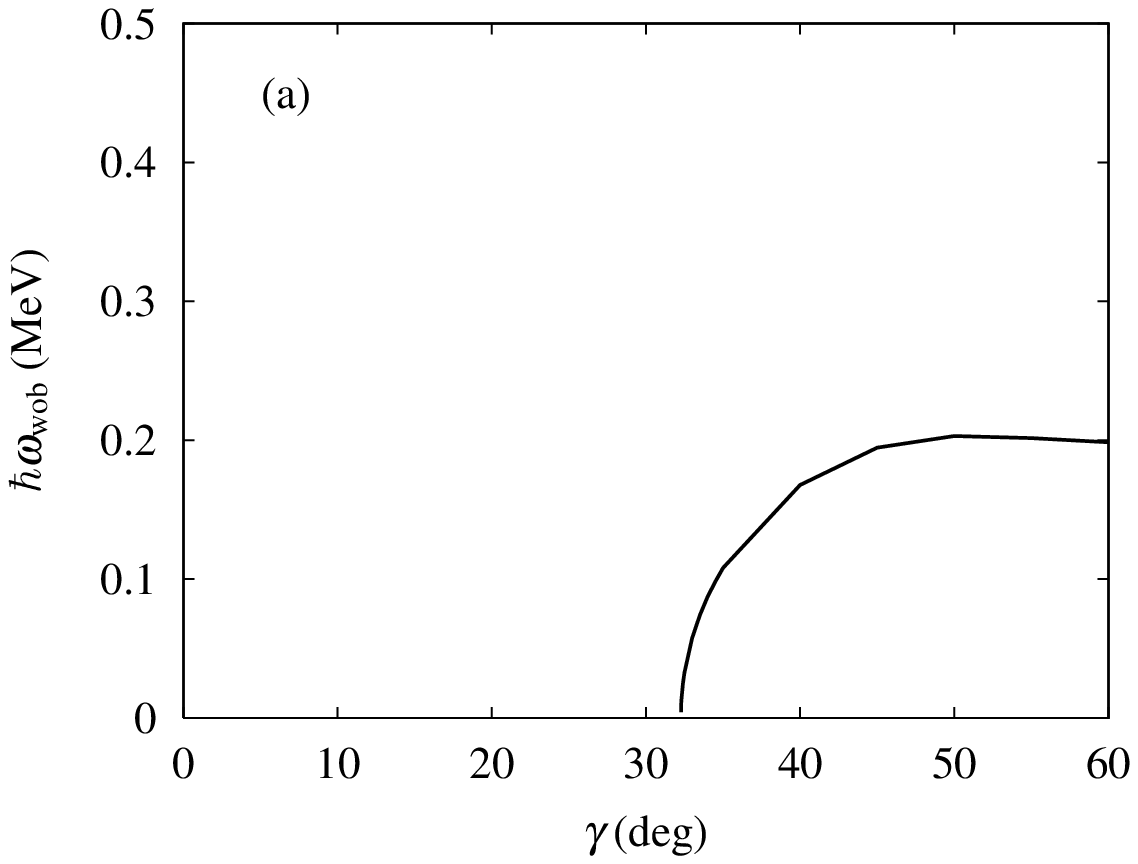}
  \includegraphics[width=7cm,keepaspectratio]{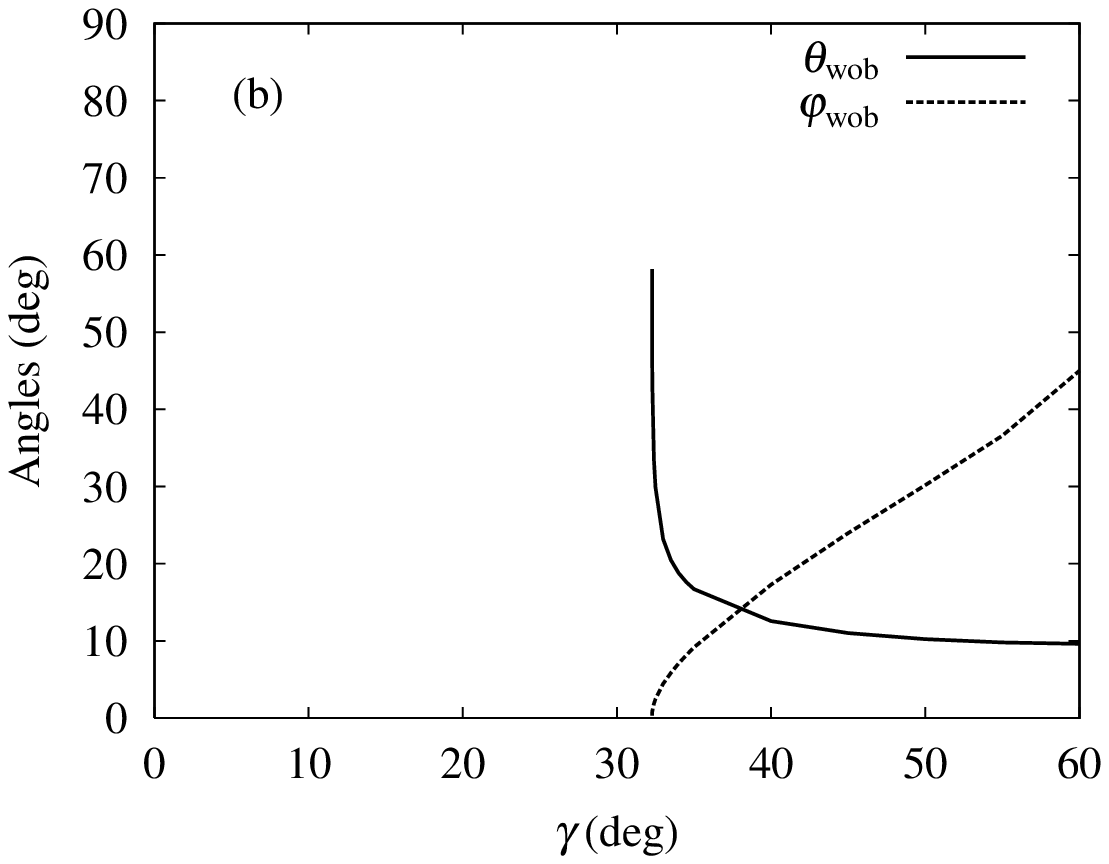}
  \includegraphics[width=7cm,keepaspectratio]{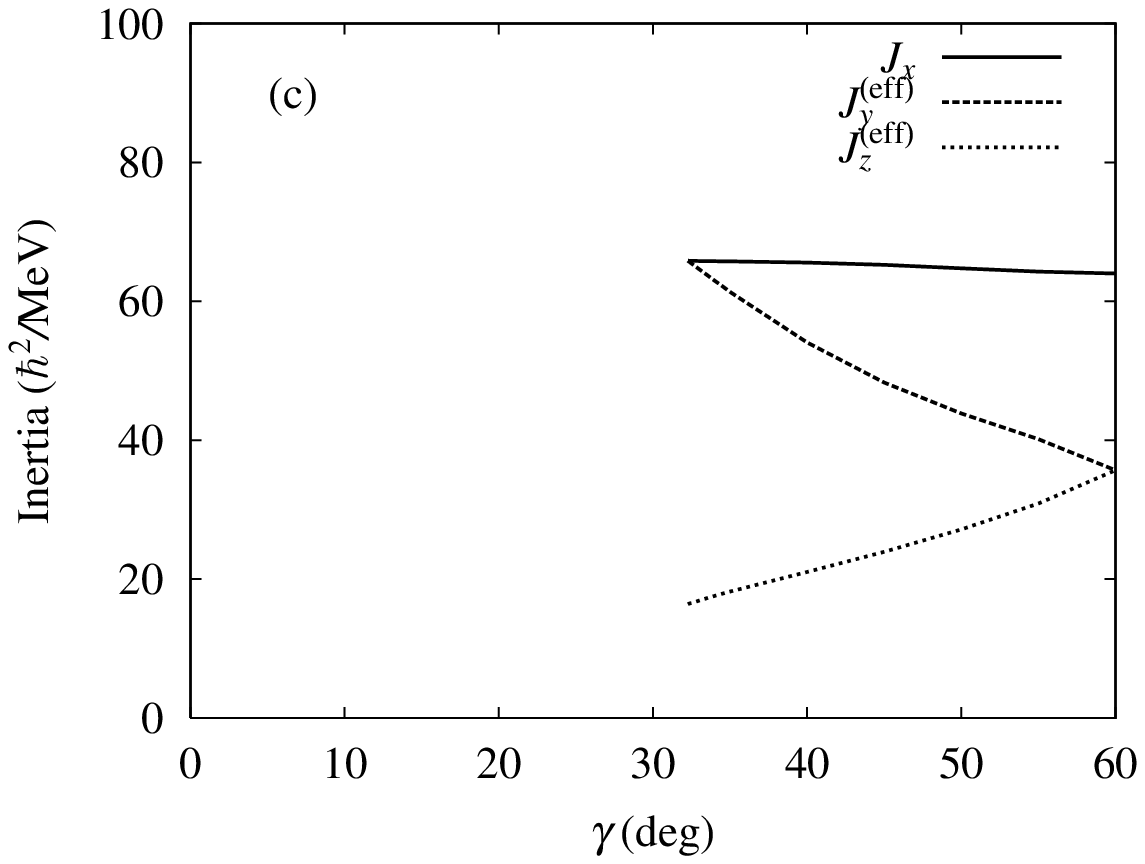}
 \caption{Triaxiality dependence of (a) excitation energy of the wobbling motion, 
          (b) wobbling angles, and (c) three moments of inertia associated with it 
          in $^{146}$Gd, 
          calculated at $\hbar\omega_\mathrm{rot} =$ 0.25 MeV with 
          $\epsilon_2 = 0.19$, $\Delta_n =$ 0.8 MeV, and  $\Delta_p =$ 0.6 MeV. 
          \label{fig1}}
\end{figure}

 In the present study we concentrate on the changes in the system with $\gamma$. 
Figure \ref{fig1}(a) reports the excitation energy $\hbar\omega_\mathrm{wob}$ 
in the rotating frame. That in the laboratory frame in the case of 
$\gamma = 60^\circ$ is given by 
$\hbar\omega_\mathrm{wob}+\hbar\omega_\mathrm{rot} =$ 0.198 MeV $+$ 0.25 MeV. 
The excitation energy decreases steeply as $\gamma$ decreases. 
In order to see its implication, we show in Fig.~\ref{fig1}(b) the wobbling angles, 
\begin{gather}
\theta_\mathrm{wob}=\tan^{-1}
{\frac{\sqrt{\vert J_y^\mathrm{(PA)}(\omega_\mathrm{wob})\vert^2
            +\vert J_z^\mathrm{(PA)}(\omega_\mathrm{wob})\vert^2}}
      {\langle J_x\rangle}} , \\
\varphi_\mathrm{wob}=\tan^{-1}
\Bigg|\frac{J_z^\mathrm{(PA)}(\omega_\mathrm{wob})}{J_y^\mathrm{(PA)}(\omega_\mathrm{wob})}\Bigg| ,
\end{gather}
with (PA) denoting the principal axis. 
$\theta_\mathrm{wob}$ clearly proves that the softening of the excitation energy 
is accompanied by a growth of the amplitude of the motion. $\varphi_\mathrm{wob}$ 
indicates that the fluctuation to the $y$ direction grows. 
Corresponding to this, the three moments of inertia behave as in Fig.~\ref{fig1}(c). 
Qualitatively, this behavior can be understood as an 
irrotational-like moments of inertia,
\begin{equation}
\mathcal{J}_k^\mathrm{irr}\propto\sin^2{(\gamma+\frac{2}{3}\pi k)} ,
\label{irr}
\end{equation}
where $k =$ 1 -- 3 denoting the $x$ -- $z$ components, superimposed by the 
contribution from the alignment, $\Delta\mathcal{J}_x$. Alternatively, it can 
also be viewed as that, at large $\gamma$, multiple alignments lead to a rigid-body-like 
inertia, 
\begin{equation}
\mathcal{J}_k^\mathrm{rig}\propto
\left(1-\sqrt{\frac{5}{4\pi}}\beta\cos{(\gamma+\frac{2}{3}\pi k)}\right) ,
\label{rig}
\end{equation}
with $\beta$ being a deformation parameter defined by the mass distribution. 

\begin{figure}[htbp]
  \includegraphics[width=5.5cm,angle=-90,keepaspectratio]{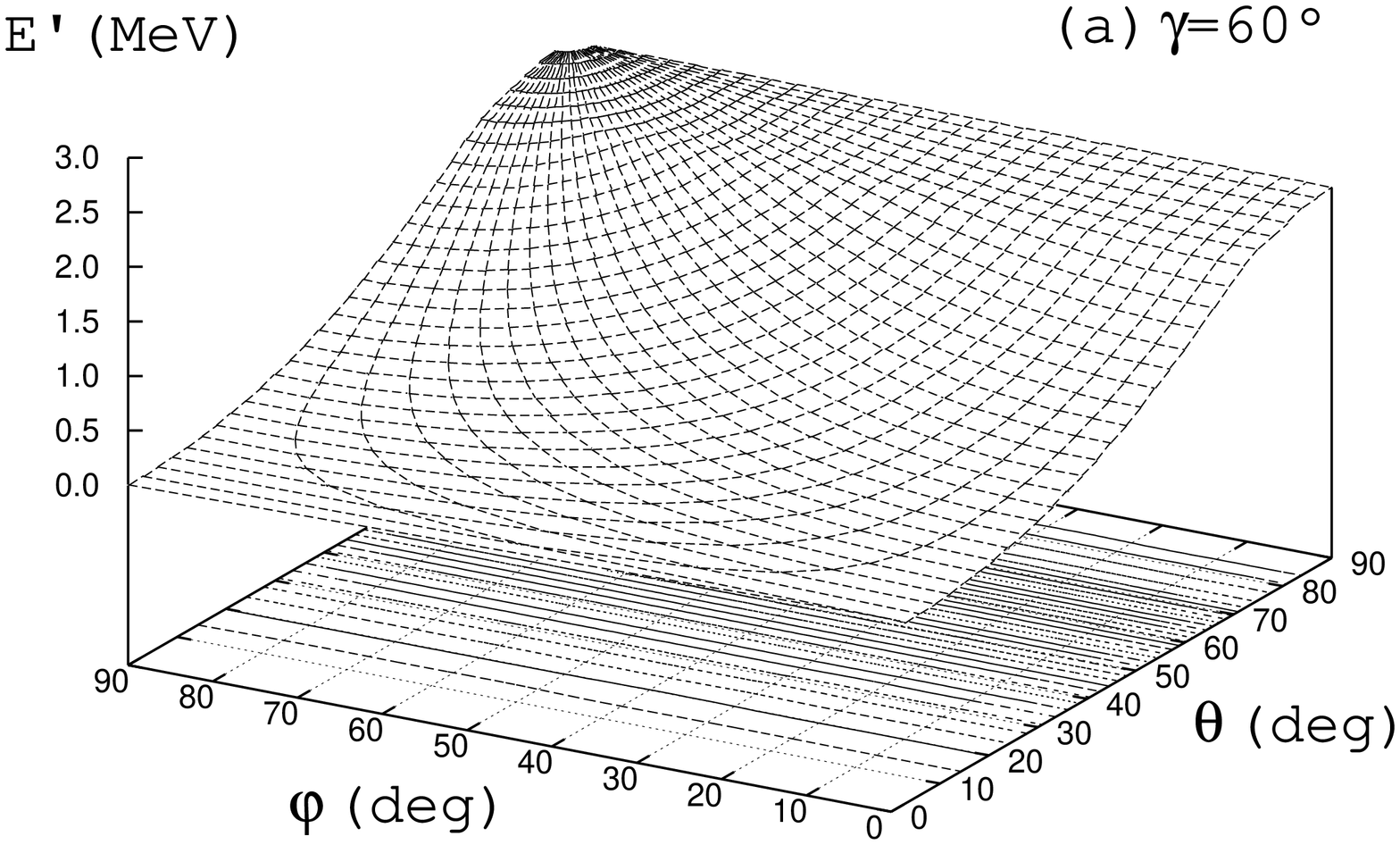}
  \includegraphics[width=5.5cm,angle=-90,keepaspectratio]{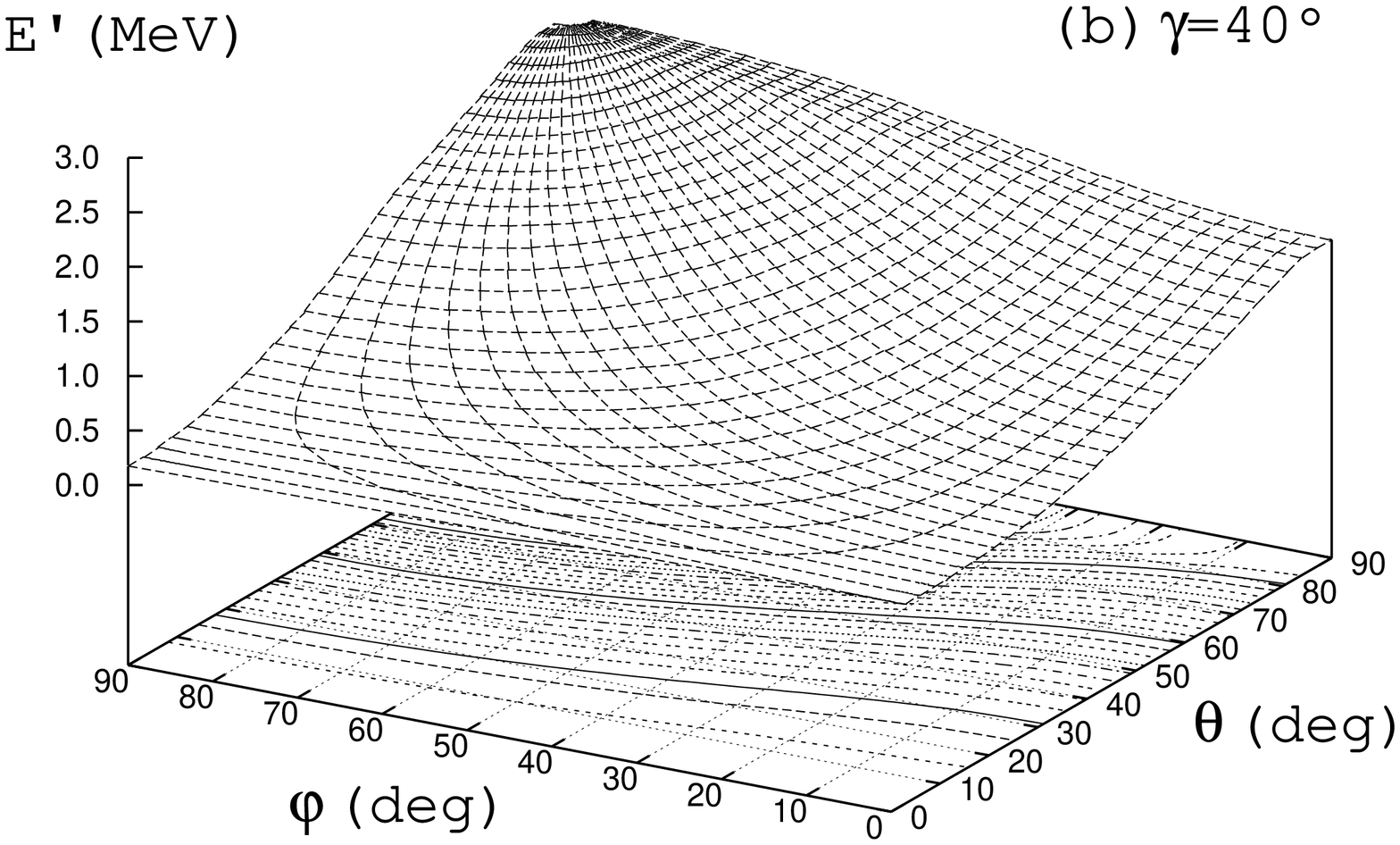}
  \includegraphics[width=5.5cm,angle=-90,keepaspectratio]{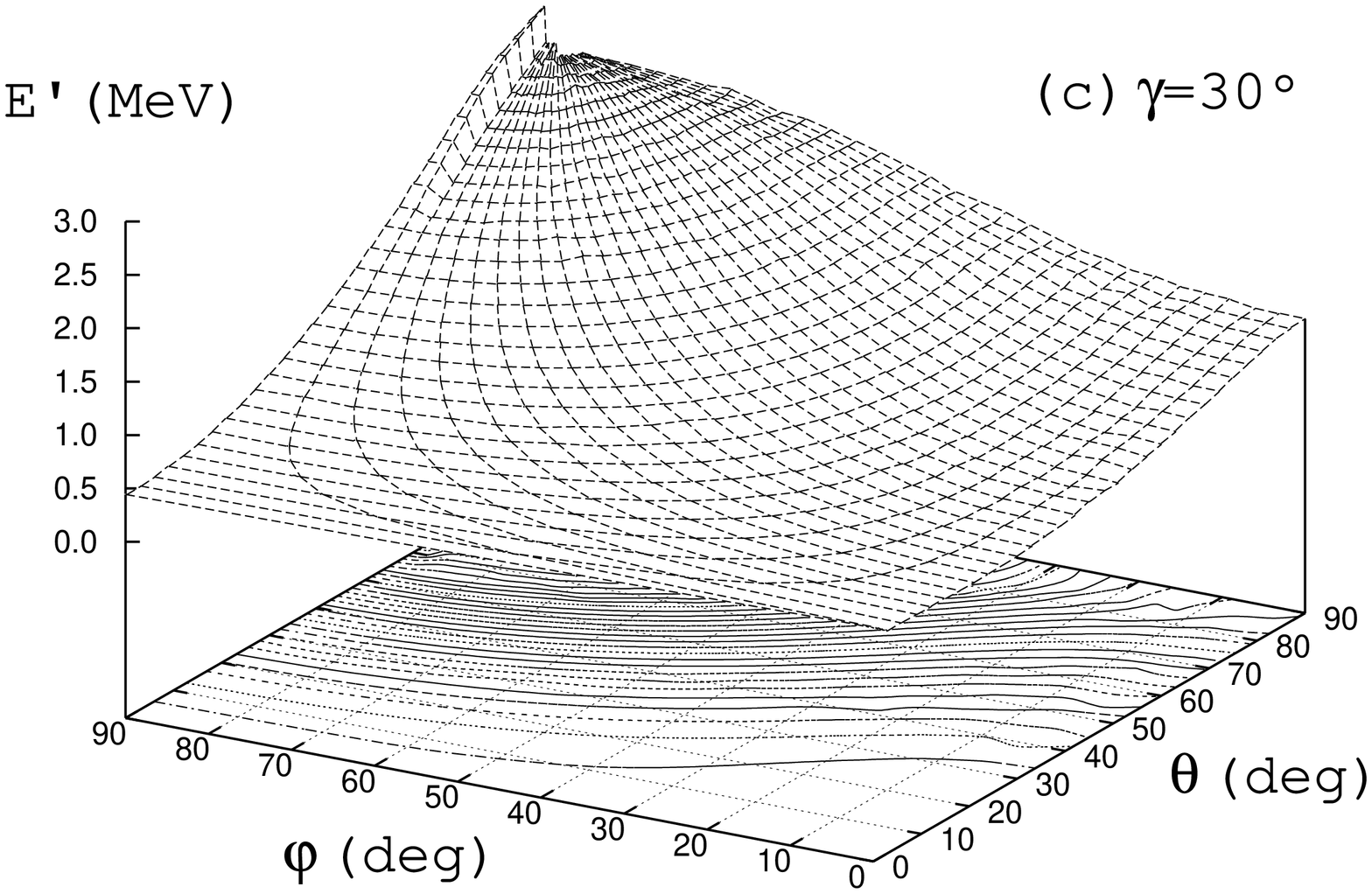}
  \includegraphics[width=5.5cm,angle=-90,keepaspectratio]{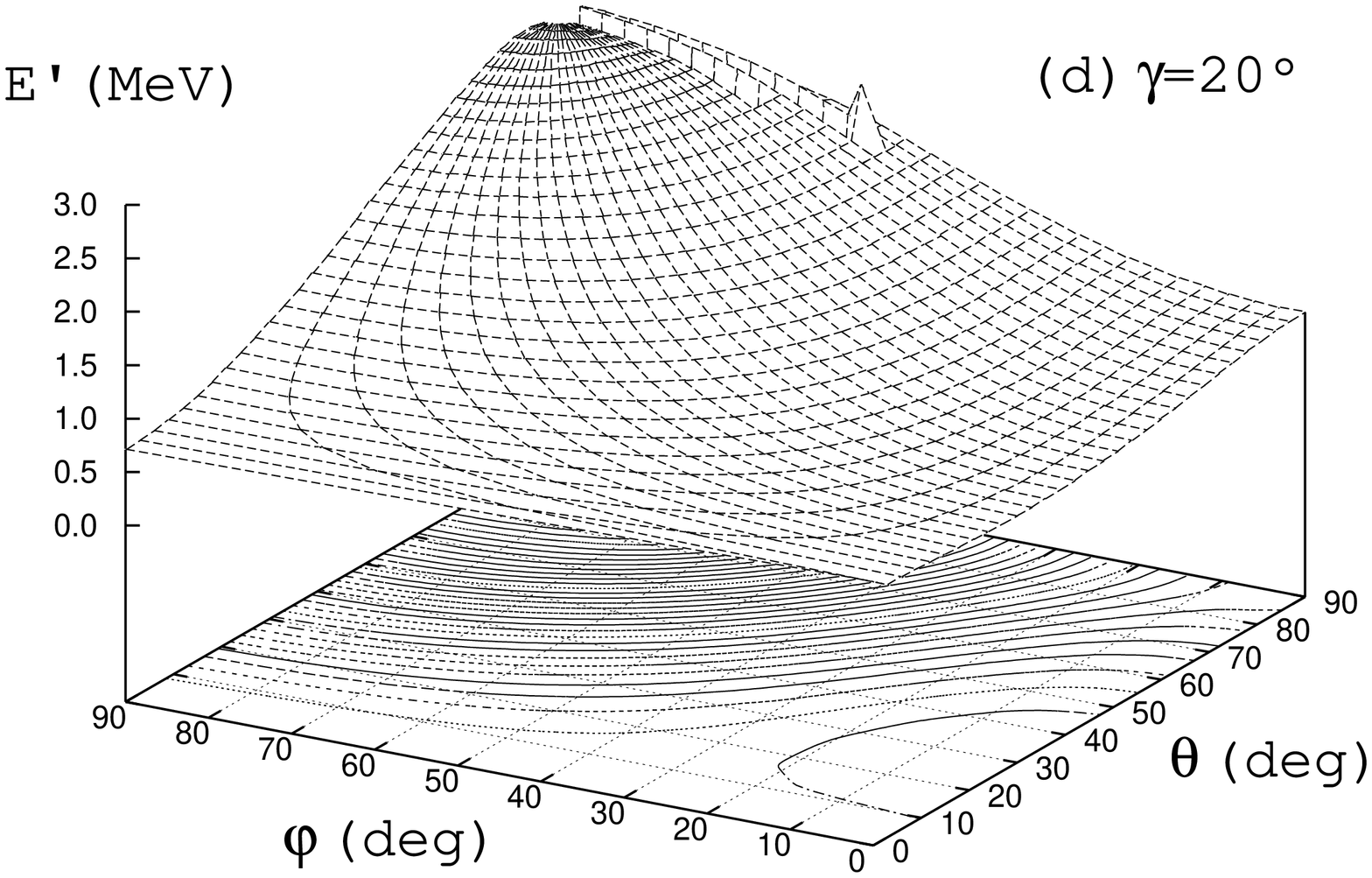}
  \includegraphics[width=5.5cm,angle=-90,keepaspectratio]{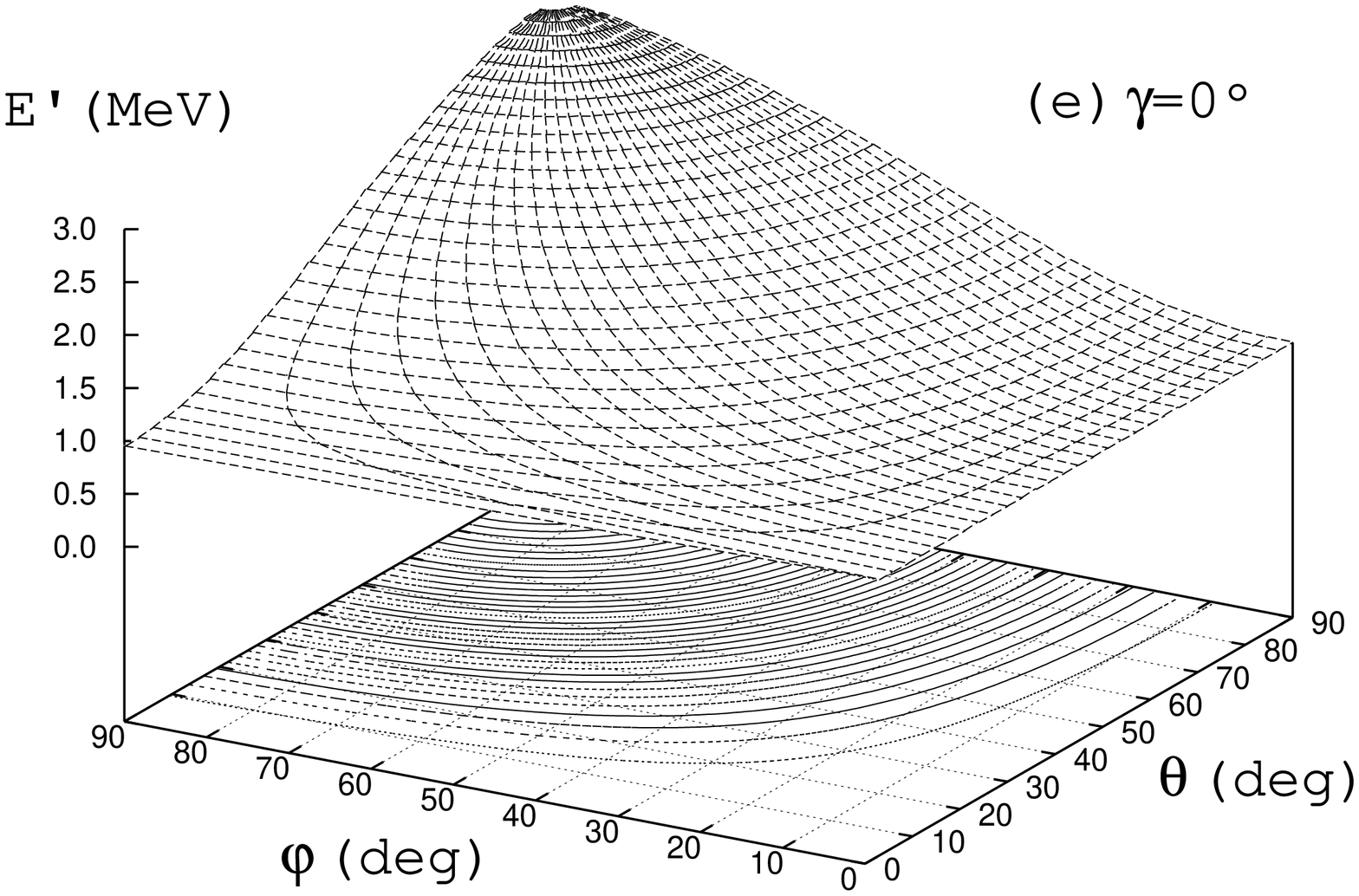}
 \caption{Energy surfaces of the 
          $[(\nu h_{9/2},f_{7/2})^2(\pi h_{11/2})^2]_{16^+}$ configuration in 
          $^{146}$Gd as functions of the tilting angle $(\theta, \varphi)$ 
          calculated with the same parameters as Fig.~\ref{fig1}, 
          (a) $\gamma = 60^\circ$, (b) $\gamma = 40^\circ$, (c) $\gamma = 30^\circ$, 
          (d) $\gamma = 20^\circ$, and (e) $\gamma = 0^\circ$.
          The interval of contours is 50 keV. Discontinuities in the surfaces are 
          due to quasiparticle crossings. 
          \label{fig2}}
\end{figure}

 Now we proceed to three-dimensional calculations; we calculate energy surfaces 
as functions of the tilting angle $(\theta, \varphi)$ of $\mathbf{\Omega}$. 
Here we note that the $(\theta, \varphi)$ plane is 
represented as a rectangle although $\varphi$ is meaningful for $\theta\neq 0^\circ$. 
Figure \ref{fig2}(a) shows the $\gamma = 60^\circ$ 
(symmetric about the $x$ axis) case. Until down to $\gamma \sim 40^\circ$, 
energy surfaces are qualitatively similar aside from becoming shallow gradually. 
But a further decrease of $\gamma$ leads to an instability of the motion to the 
$\theta$ direction with $\varphi = 0^\circ$, that is, the direction of the $y$ axis. 
Together with the 
property that the surface is stable with respect to the direction of the $z$ axis, 
the situation corresponds excellently to Fig.~\ref{fig1}. The behavior of 
$\varphi_\mathrm{wob}$ in Fig.~\ref{fig1}(b) can be interpreted as that, when the system can 
fluctuate to the direction of the $y$ axis without any energy cost, it does not 
fluctuate to the $z$ axis. 

\begin{figure}[htbp]
  \includegraphics[width=7cm,keepaspectratio]{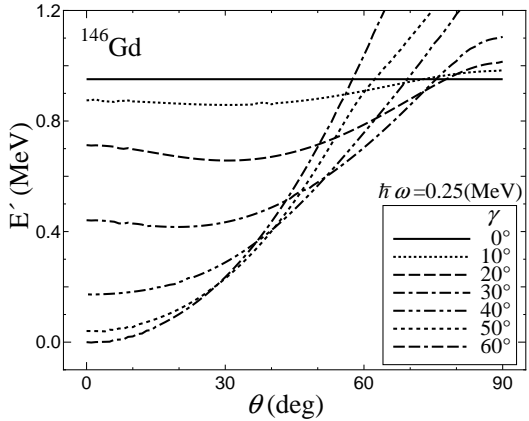}
 \caption{Cross sections at $\varphi = 0^\circ$ of the energy surfaces of $^{146}$Gd.
          \label{fig3}}
\end{figure}

 To look at the energy surface more closely we gather their cross sections at 
$\varphi = 0^\circ$ (the $x$-$y$ plane) in Fig.~\ref{fig3}. This figure clearly shows 
that a tilted axis minimum appears at around $\gamma = 30^\circ$ although it is 
shallow. The correspondence to Fig.~\ref{fig1} in which the instability occurs 
at $\gamma = 32^\circ$ is excellent. Note that the 
reason why the wobbling angle seen from Fig.~\ref{fig3} is larger than $\theta_\mathrm{wob}$ 
in Fig.~\ref{fig1}(b) is that this is drawn for $\mathbf{\Omega}$.

\section{Potential surface for the wobbling mode in $^{163}$Lu}

\begin{figure}[htbp]
  \includegraphics[width=8.5cm,keepaspectratio]{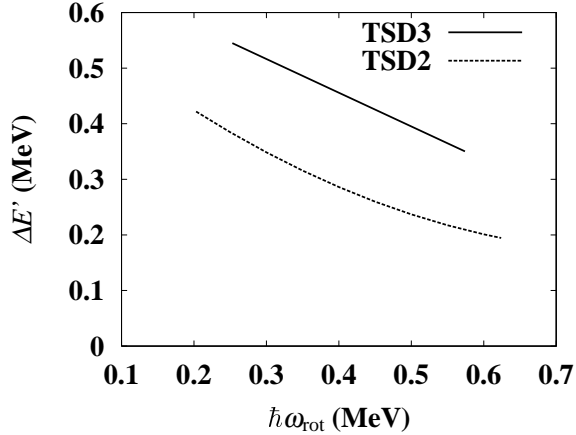}
 \caption{Experimental excitation energies of the two- and one-phonon wobbling 
          states relative to the yrast triaxial superdeformed states in 
          $^{163}$Lu. Data are taken from Ref.~\cite{2phonon}.
          \label{fig4}}
\end{figure}

\begin{figure}[htbp]
  \includegraphics[width=7cm,angle=-90,keepaspectratio]{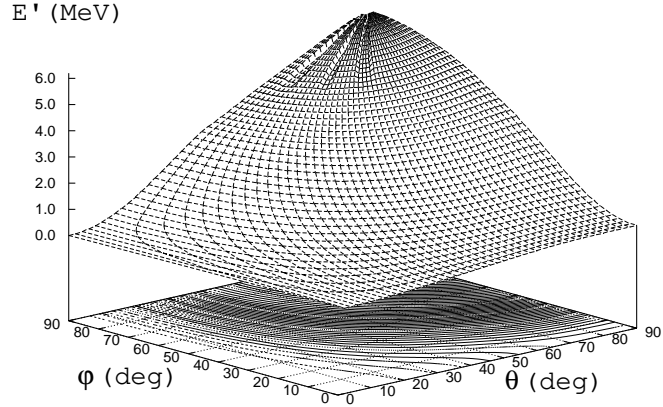}
 \caption{Energy surface of the triaxial superdeformed one-quasiparticle configuration in 
          $^{163}$Lu as a function of the tilting angle $(\theta, \varphi)$
          calculated at $\hbar\omega_\mathrm{rot} =$ 0.5 MeV with 
          $\epsilon_2=0.43$, $\gamma=20^\circ$, and $\Delta_n=\Delta_p= 0.3$ MeV. 
          The interval of contours is 100 keV. 
          \label{fig5}}
\end{figure}

 The analyses above are purely theoretical. Then, is there any experimental 
signature of the softening of the wobbling motion? We think the answer is yes. 
Figure \ref{fig4} shows the experimental~\cite{2phonon} excitation energies 
(in the rotating frame) of the TSD3 (two-phonon wobbling) and the TSD2 
(one-phonon wobbling) relative to the TSD1 (yrast 1QP TSD) in $^{163}$Lu, where TSD is 
the abbreviation for triaxial superdeformation. 
$\Delta E'_\mathrm{2-phonon} < 2\times\Delta E'_\mathrm{1-phonon}$ indicates a 
signature of softening of the energy surface. 
We obtained $\hbar\omega_\mathrm{wob} = 0.185$ MeV, $\theta_\mathrm{wob} = 14.2^\circ$, 
and $\varphi_\mathrm{wob} = 7.6^\circ$ for the one-phonon wobbling state in the RPA 
(see also Refs.\cite{msmr,msmr2} for the RPA calculation). 
The small value of $\varphi_\mathrm{wob}$ looks to indicate a softening to the $y$ direction. 
Calculated energy surface is shown in Fig.~\ref{fig5}. Calculations were done 
in the model space of five major shells, $N_\mathrm{osc} =$ 3 -- 7 for neutrons 
and 2 -- 6 for protons, with 
$\epsilon_2=0.43$, $\gamma=20^\circ$, $\Delta_n=\Delta_p= 0.3$ MeV, and 
$\hbar\omega_\mathrm{rot} = 0.5$ MeV where the calculated $\hbar\omega_\mathrm{wob}$ 
approaches the experimental one. This figure shows again the surface 
softens to the direction of the $y$ axis. 

\begin{figure}[htbp]
  \includegraphics[width=7cm,angle=-90,keepaspectratio]{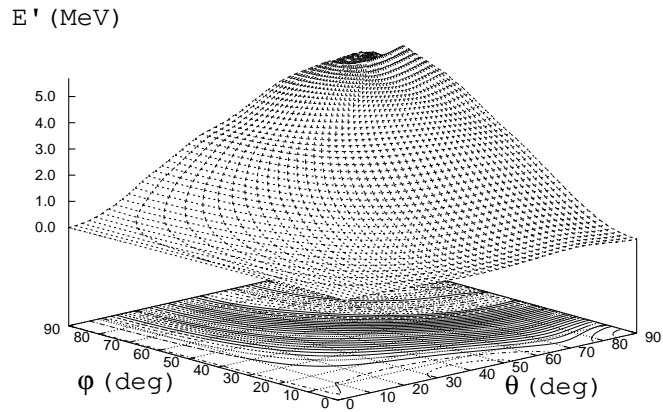}
 \caption{The same as Fig.~\ref{fig5} but for the zero-quasiparticle configuration 
          in $^{162}$Yb. 
          \label{fig6}}
\end{figure}

 In Refs.\cite{msmr,msmr2}, it was shown that the alignment of the $\pi i_{13/2}$ 
quasiparticle was essential for the appearance of the wobbling motion. In order to 
see this fact from the viewpoint of the potential surface, we calculated the 0QP 
(non-yrast TSD at high spins) configuration in $^{162}$Yb. Figure~\ref{fig6} clearly 
shows that a tilted axis minimum in the $x$-$y$ plane is realized when the wobbling 
motion does not occur due to the lack of $\pi i_{13/2}$ QPs that make $\mathcal{J}_x$ 
larger than $\mathcal{J}_y^\mathrm{(eff)}$~\cite{msmr,msmr2}. 
This result proves that the low-$\Omega$ 
high-$j$ orbital favors the principal axis rotation on which the wobbling motion occurs. 

 For deeper understanding of the two-phonon states, application of 
more sophisticated many body theories such as the selfconsistent collective coordinate 
(SCC) method~\cite{scc} is desirable. 

\section{Summary}

 To summarize, we have proved that a tilted axis rotation emerges when the wobbling mode 
becomes unstable as the triaxiality parameter changes in an oblate configuration in 
$^{146}$Gd. Its instability is caused by a growth of the fluctuation of the motion 
of the angular momentum or frequency vector to the direction of the $y$ axis. 
Having performed this theoretical calculation, we have argued that the signature of 
the softening of the wobbling motion can be seen in the observed spectra of the triaxial 
superdeformation in $^{163}$Lu and shown that a tilted axis minimum would appear if it 
were not for the $\pi i_{13/2}$ quasiparticle. 

\begin{acknowledgments}
 The numerical calculations in this work were performed in part
with the computer system of the Yukawa Institute Computer
Facility, Kyoto University.
\end{acknowledgments}

\end{document}